\documentclass[11pt,titlepage]{article}

\usepackage{setspace} 

\usepackage{amsmath}

\usepackage{graphicx}
\usepackage{amssymb}
\usepackage{epstopdf}

\usepackage[round,numbers,sort&compress]{natbib} 
\newcommand{\vc}{{\bf c}}

\newcommand{\n}{{\bf n}}
\newcommand{\F}{{\bf F}}

\title{Cross-talk and interference enhance information capacity of a signaling pathway}

\author{Sahand~Hormoz\thanks{
          Corresponding author.  Address: 
          Kavli Institute for Theoretical Physics,
	   Kohn Hall,
	   University of California, 
	   Santa Barbara, CA~93106, U.S.A.} \\
	Kavli Institute for Theoretical Physics, \\
	University of California Santa Barbara}

\date{}

\begin{document}

\maketitle

\abstract{A recurring motif in gene regulatory networks is transcription factors (TFs) that regulate each other, and then bind to overlapping sites on DNA, where they interact and synergistically control transcription of a target gene. Here, we suggest that this motif maximizes information flow in a noisy network. Gene expression is an inherently noisy process due to thermal fluctuations and the small number of molecules involved.  A consequence of multiple TFs interacting at overlapping binding-sites is that their binding noise becomes correlated. Using concepts from information theory, we show that in general a signaling pathway transmits more information if 1) noise of one input is correlated with that of the other, 2) input signals are not chosen independently. In the case of TFs, the latter criterion hints at up-stream cross-regulation. We demonstrate these ideas for competing TFs and feed-forward gene regulatory modules, and discuss generalizations to other signaling pathways. Our results challenge the conventional approach of treating biological noise as uncorrelated fluctuations, and present a systematic method for understanding TF cross-regulation networks either from direct measurements of binding noise, or bioinformatic analysis of overlapping binding-sites.

}

\clearpage

\section*{Introduction}

Acurate transmission of information is of paramount importance in biology. For example in the process of embryonic development, crude morphogen gradients need to be translated into precise expression levels in every cell and sharp boundaries between adjacent ones  \citep{Wolpert1,Wolpert2}. The embryo accomplishes this using a complex network of signaling molecules that not only regulate the expression level of the desired output gene, but also each other. One simple strategy for increasing accuracy is the use of multiple input signals. Indeed, frequently, the expression level of a single gene is controlled by multiple transcription factors (take for example {\it bicoid} and {\it hunchback}, or {\it dorsal} and {\it twist} in the {\it Drosophila} embryo \citep{Wolpert1,Levine1,Levine2}). These transcription factors, however, often have overlapping binding sites that result in interactions at binding and synergetic control of transcription \citep{Levine1,Levine2}.

Here, we suggest that interaction at the level of binding (interference) is related to the upstream network of transcription factors regulating each other (cross-talk). Our main assumption is that the regulatory network is designed to optimize information transfer from the input (TF concentrations) to the output (gene expression level). This is a reasonable assumption in the case of development, where accurate positional information needs to be extracted from noisy morphogen concentrations \citep{Wolpert2}. 

First, we define the concept of a cis-regulatory network as a noisy communication channel, where the input encodes information by taking on a range of values, i.e. a morphogen gradient that carries positional information. Decoding this information is subject to biological noise; for example, at the molecular level, the stochastic binding of morphogens to receptors makes an exact read-out of their concentration impossible. 

We show that in general two input signals with correlated noise can transmit more information if they are not independent information carriers but chosen from an `entangled' joint-distribution, i.e. the concentration of one morphogen in a given cell is related to concentration of the other. Physically, this implies that the two inputs regulate each other upstream through cross-talk. We demonstrate this by analyzing a simple model of two TFs competing for the same binding site. The competition at the binding site results in correlated binding/unbinding fluctuations. Solving for the optimal joint-distribution of the input TF concentrations indicates that upstream, one is positively regulated by the other. Despite the increase in noise for each individual input from the competition, two interacting TFs can transmit more information than two non-interacting TFs because of 1) correlated noise in the inputs, 2) an entangled optimal input distribution.

We suggest that this mechanism is consistent with the recurring strategy of the feed-forward motif, where one TF positively regulates another, and both bind to partially overlapping sites that induce interactions. We confirm this claim by simulating the stochastic dynamics of all structural types of the feed-forward loop subject to correlated input noise. Three specific biological examples are discussed: joint regulation of gene {\it Race} in the {\it Drosophila} embryo by intracellular protein Smads and its target zen; regulation of {\it even-skipped} stripe 2 by {\it bicoid} and {\it hunchback}; and that of {\it snail} by {\it dorsal} and {\it twist}. Generalization to other forms of cross-talk, such as cross-phospohorylation, and other forms of interference, such as use of scaffold proteins, are also discussed.

The structure of this paper is as follows: in the Results section, we first establish that in general two input signals with correlated noise can transmit more information if their input joint-distribution is not separable. Next, we consider a model of two TFs competing for the same binding site. We calculate analytically the noise correlations and the associated optimal input distribution. In some regimes, competing TFs outperform independent ones. Motivated by this finding, in the last subsection of the Results, we ask whether realistic gene regulatory modules can generate close to optimal input distributions and combine correlated inputs to maximize information transmission. We compute numerically the channel capacity for fee-forward loops, where the joint-regulation of the target gene is subject to generic correlated noise; noise correlations are a parameter in these simulations to generalize the results beyond competing TFs. Relevant biological examples are considered in the Discussion.

\subsection*{Gene regulation as a communication channel}

Regulatory networks in a cell are information processing modules that take in an input, such as concentration of a nutrient, and generate an output in the form of a gene expression level. Information in the input is typically encoded as the steady-state concentration of a transcription factor $c$, which binds to the promoter site of the desired response gene and enhances or inhibits its transcription. At a molecular level, the process of binding is inherently noisy, subject to thermal agitations and low-copy number fluctuations \citep{Elowitz1,Elowitz2,Paulsson}. The noise is captured through a probabilistic relationship between the TF concentration $c$, and gene expression level $g$, $P(g|c)$. Detailed form of $P(g|c)$ depends on physical parameters such as binding and unbinding rates. We can think of this process as communication across a noisy channel \cite{WalczakReview}. To alleviate impact of the noise, various strategies can be adopted, such as limiting the input to sufficiently spaced discrete concentration levels $c_i$ that result in non-overlapping outputs. In many gene regulatory networks, spatial and temporal averaging of input signals are also used to reduce noise \citep{Gregor}.

Shannon's channel coding theorem \citep{Shannon,Cover} tells us the maximum rate at which information can be communicated across a noisy channel, or the channel capacity. Throughout this work, we will assume that gene regulatory networks are selected to optimize the rate of information transmission. This is a strong but reasonable assumption; for example, the cell will clearly benefit from a more accurate knowledge of the amount of nutrient in its environment. However, the cost of an optimal networks can exceed the benefit of more accurate information. Here, we do not account for the cost of a network, the only metric for comparison is the channel capacity.

With knowledge of the nature of the noise in a channel $P(g|c)$, it is possible to compute the probability distribution of the input signal $P_{TF}^*(c)$ that maximizes rate of information transmission. Essentially, this distribution tells the sender how often a particular TF concentration should be used for optimal transmission of information encoded in concentration. However, it does not tell the sender anything about the encoding and decoding schemes. This abstraction is useful, allowing us to compute the optimal input without having derived the optimal coding. However, the optimal coding might require input blocks of infinite size and complex codebooks, with little biological relevance. 

Nevertheless, there are experimental observations consistent with the idea of regulatory systems maximizing information transmission rates. Tkacik et al. \citep{BialekPNAS} have shown that experimental measurements of Hunchback concentration in early Drosophila embryo cells \citep{Gregor} has a distribution that closely matches the optimal frequency for the measured levels of noise in the system; with the system achieving ~90\% of its maximum transmission rate.

\section*{Methods}

\subsection*{Competing TFs binding model}

The fractional occupation of the binding site by TF $i$ ($n_i$) satisfies the kinetic equation 
\begin{equation}
\frac{dn_i(t)}{dt} = k c_i \big( 1 - n_1 - n_2 \big) - l n_i + \xi_{n_i} \ \ \ i=1,2
\end{equation}
where $k$, $c_i$ and $l$ are the on-rate, TF concentration, and off-rate respectively. The Langevin noise term $\xi_{n_i}$ introduces uncorrelated fluctuations: $\langle \xi_{n_i} (t)\rangle = 0$ and $\langle \xi_{n_i}(t) \xi_{n_j}(t') \rangle = D_i^2 \delta_{ij} \delta (t' - t) $  . For independent TFs, the first term on the right hand side is modified to $k c_i \big( 1 - n_i \big)$. For Fig.4A and B, above stochastic differential equations were numerically integrated using Euler-Maruyama method \citep{Higham} discretized with dt=0.01, for parameters, $k = 1$, $c_{1,2} = 0.01$, $l = 10^{-4}$, and $D^2_i = k c_i \big( 1 - \bar{n}_1 - \bar{n}_2 \big) + l \bar{n}_i = 2 l \bar{n}_i$. $\bar{n}_i$ denotes the average steady state value of $n_i$. In Fig. 4A, the power spectral density is computed using the Wiener-Khinchin formula $S_{i}(\omega) = \frac{1}{T} \big\langle \left| \int_0^T n_i(t) e^{-i \omega t} dt \right|^2 \big\rangle$ \cite{Papoulis}. 

\subsection*{Feed-forward motif kinetic equations} 
The concentrations of $Y$ and $g$ are given by the kinetic equations \citep{Alon1,Alon2,Tyson}
\begin{eqnarray*}
dY/dt &=& \beta_y f(X + \zeta_x,K_{xy}) - \alpha_y Y  + \eta_y\\
dg/dt &=& \beta_g F(X + \xi_x , K_{xg} ; Y + \xi_y , K_{yg}) - \alpha_g g + \eta_g.
\end{eqnarray*}
$f$ can be an activator, $f(x,k) = (x/k)^H/(1+(x/k)^H)$, or repressor, $f(x,k) = 1/(1+(x/k)^H)$, where $H$ is the Hill coefficient. $K_{ij}$ is the regulation coefficient of gene j by TF i. AND-gate: $F(x,k_x;y,k_y) = f(x,k_x)f(y,k_y)$. OR-gate: $F = f_o(x;k_x,y,k_y)+f_o(y;k_y,x,k_x)$, where for an activator $f_o(x;k_x,y,k_y) = (x/k_x)^H/(1+(x/k_x)^H+(y/k_y)^H)$, and a repressor $f_o(x;k_x,y,k_y) = 1/(1+(x/k_x)^H+(y/k_y)^H)$.

The top equation captures cross-regulation of TF $Y$ by TF $X$. The output noise is captured by the Langevin  term $\eta_y$. The input noise --from fluctuations in the read-out of the TF $X$ concentration due to binding fluctuations, diffusion noise, etc. -- is captured by the phenomenological noise term $\zeta_x$. The noise in cross-regulations is an extrinsic noise in the system, since by definition our channel is defined as joint-regulation of gene $g$ by TFs $X$ and $Y$ (bottom equation).

The intrinsic noise contains the output noise in synthesis and degradation fluctuations --shot noise-- in $g$ captured by Langevin term $\eta_g$. The intrinsic input noise is due to fluctuations in the read-outs of the TF concentrations $X$ and $Y$ which determine the synthesis rate of $g$ through the Hill function $F$. The inputs of $F$ fluctuate from the true concentration of $X$ and $Y$ by stochastic terms $\xi_x$ and $\xi_y$ respectively.

\subsection*{Phenomenological Noise} 
We neglect the contribution of extrinsic noise; keeping it does not change the results qualitatively --namely, the connection between correlated input noise and upstream cross-regulation. The intrinsic input noise in TF concentration read-out satisfy $\langle \xi_{x,y} (t) \rangle = 0$, and a general phenomenological form for their variance set by a constant term and one proportional to the TF concentration, since regardless of microscopic details, this noise stems fundamentally from finite, discrete, and fluctuating molecule numbers: $\langle \xi_x(t) \xi_x(t') \rangle = \big(\epsilon + qX\big)\delta(t'-t)$, $\langle \xi_y(t) \xi_y(t') \rangle = \big( \epsilon + qY \big) \delta(t'-t)$. The input noise can be correlated: $\langle \xi_x(t) \xi_y(t') \rangle = \rho q\sqrt{XY} \delta(t'-t)$; $\rho$ is the noise correlation coefficient, which is assumed to be independent of the TF concentrations for simplicity. A more complex structure for $\rho$ --for instance with concentration dependence as in the case of competing TFs-- does not change the results qualitatively. $\epsilon$ is a small constant that ensures a minimum noise of one TF molecule per cell. For the output noise, $\langle \eta_g(t) \rangle = 0$ and $\langle \eta_g(t)\eta_g(t') \rangle = \big(\epsilon + qg\big) \delta(t'-t)$. 

\subsection*{Numerical simulations}
Initial conditions: at time t=0, Y=0 and g=0. X=$c_1$ for all times t$\ge$0. Above stochastic differential equations were numerically integrated using Euler-Maruyama method \citep{Higham} from t=0 to t=10 discretized with dt=0.001. Output statistics were gathered after steady state is reached, last 3000 time steps, for 1000 runs. Implemented in MATLAB R2011a.

\subsection*{Input distribution optimization}
Output distribution $P(g|c_1)$ was computed for 30 values of $c_1$ equally spaced in the log scale from $log(c_1) = -1$ to $log(c_1)=1.5$ (see Model parameters below). Discretization captures spatial averaging of the diffusing inputs (exponentially decaying from source) by the cells \citep{Gregor2}. Using a higher resolution over the input range did not increase the capacity significantly since optimizing over input distribution resulted in discrete (spaced-out) inputs. Constrained non-linear optimization with a sequential quadratic programming (SQP) method \citep{Powell} was used to numerically optimize over the input distribution $P_{in}(c_1)$ and compute channel capacity. Implemented in MATLAB R2011a.

\subsection*{Model parameters}
The range of input and noise parameters were selected to match that of experimental measurements of morphogen Hunchback in Drosophila embryo \citep{Gregor}. The conclusions above were unaffected by changing the parameters as long as the FFLs had dynamics with non-trivial steady states. However, for the figures and numbers quoted in the text the following parameters were used: $\alpha_g$ = $\beta_g$ = 100, $\beta_y$ = 10, $\alpha_y$ = 1, $K_{xy}$ = $K_{xg}$ = $K_{yg}$ = 1, H = 1, q = 1, $\epsilon$ = 0.001.

\section*{Results}

\subsection*{Noise correlations enhance capacity}

First, we quantify how correlations in noise of multiple inputs enhance rate of information transmission, following closely the approach of \citep{BialekPRE1}. Consider two transcription factors with concentrations $c_1$ and $c_2$ that regulate expression level of a gene (denoted as $g$). These values can vary for example as a function of space, as in the case of morphogens along an embryo. The frequency of observing a particular concentration occurrence $c_1$ and $c_2$ is given by $P_{TF}(c_1,c_2)$. The entropy --or uncertainty-- of the inputs is maximized when this distribution is uniform, or all concentrations equally likely, which implies that the maximum amount of information is gained when the TF concentrations are determined precisely. Of course, our aim is not to maximize the entropy in $c_{1,2}$, but rather the information conveyed to the expression level $g$.

The noise in the expression levels results in a distribution of $g$ for fixed TF concentrations, $P(g|c_1,c_2)$. Equivalently, we can fix the expression level $g$ and consider the corresponding distribution of TFs, $P(c_1,c_2|g)$, assuming that there is unique set of inputs for every value of $g$. The two distributions are related by Bayes' rule. The amount of information communicated from $c_{1,2}$ to $g$ is given by the mutual information between the distributions of $c_{1,2}$ and $g$ \citep{Shannon},
\begin{eqnarray}
I(g;c_1,c_2) = -\int dc_1dc_2 P_{TF}(c_1,c_2) \log P_{TF}(c_1,c_2) \nonumber \\
+ \int dg P_{exp}(g) \times \int dc_1dc_2 P(c_1,c_2|g) \log P(c_1,c_2|g), \label{MutInfo}
\end{eqnarray}
where the distribution of expression level $g$ is given by $P_{exp}(g) = \int dc_1 dc_2 P(g|c_1,c_2)P_{TF}(c_1,c_2)$.

We assume that the noise in $c_{1,2}$ for a fixed expression level $g$ is small and distributed as a Gaussian around the mean value $\bar{\vc}(g)$,
\begin{equation}
P(c_1,c_2|g) = \frac{1}{2\pi \sqrt{|\Sigma|}} \exp \bigg[ -\frac{1}{2} \big( \vc - \bar{\vc}(g) \big)^T \Sigma^{-1} \big(\vc - \bar{\vc}(g) \big) \bigg], \label{NormDist}
\end{equation}
where $\vc = (c_1,c_2)$, and $\Sigma$ is the covariance matrix over the conditional probability for fixed $g$, or the noise covariance matrix, $\Sigma_{ij}(g) = \langle \big( c_i - \bar{c}_i(g) \big) \big( c_j - \bar{c}_j(g) \big) \rangle$.

The small-noise approximation says that it is meaningful to think of a mean one-to-one input-output response, which is what is commonly measured in experiments. We expand around the mean response to the next order. The approximation, although strong, has been verified for a variety of regulatory systems (see for example Bicoid-Hunchback in \citep{Gregor,Little}, or for other examples \citep{O'Shea,Newman,Rosenfeld}), and enables us to analytically calculate the optimal distribution. We will relax these assumptions later with a numerical approach. The mutual information under this approximation is given by,
\begin{eqnarray}
I(g;c_1,c_2) = -\int dc_1dc_2 P_{TF}(c_1,c_2) \log P_{TF}(c_1,c_2)  \nonumber \\
+ \frac{1}{2} \int dc_1 dc_2 P_{TF}(c_1,c_2) \log \bigg( \frac{ |\Sigma^{-1}(\bar{g}(\vc))| }{4\pi^2 e^2}\bigg) , \label{MutInfoApprox}
\end{eqnarray}
where $\Sigma^{-1}$ is evaluated at the mean value of expression level $\bar{g}$ corresponding to a given $\vc$.

To find the channel capacity, Eq.\ref{MutInfo} is optimized for the input distribution $P_{TF}(c_1,c_2)$. With the probability distribution's normalization constraint introduced using a Lagrange multiplier, the optimal distribution must satisfy,
\begin{equation}
\frac{\delta}{\delta P_{TF}(c_1,c_2)} \bigg[ I(g;c_1,c_2) - \lambda \int dc_1 dc_2 P_{TF}(c_1,c_2) \bigg] = 0.
\end{equation}
The optimal input distribution in the small-noise approximation (Eq.\ref{MutInfoApprox}) is given by,
\begin{equation}
P^*_{TF}(c_1,c_2) = \frac{1}{2\pi e Z}\frac{1}{\sqrt{|\Sigma|}} \label{optDist},
\end{equation}
where $Z$ is the normalization constant.

The maximum mutual information, or channel capacity for transmitting information from TF concentrations to expression level equals,
\begin{equation}
I^* = \log_2 Z = \log_2 \bigg[ \frac{1}{2\pi e } \int \int_{c_{min}}^1 dc_1 dc_2 \frac{1}{\sqrt{|\Sigma|}} \bigg]. \label{chanCap}
\end{equation}
We have constrained the input concentration to lie in the normalized range $c_{1,2} \in [c_{min},c_{max}=1]$. The minimum concentration is set by the molecular nature of the input: a minimum of one input molecule per cell is required.

We can repeat the same calculation for one TF while neglecting the other, effectively ignoring the covariance of the noise (off-diagonal components of $\Sigma$). With no covariance, the noise distribution is separable, $P(c_1,c_2|g)=P(c_1|g)P(c_2|g)$. The optimal input concentration for TF$_1$ will be $P^*_1(c_1) \sim \frac{1}{\sqrt{\Sigma_{11}}}$, and its channel capacity, $I^*_1 \sim \log \int dc_1 \frac{1}{\sqrt{\Sigma_{11}}}$; with a similar expression for the other TF.

For the simple case where $\Sigma$ is independent of $\vc$, channel capacity of the two TFs can be decomposed into its individual and joint contributions, 
\begin{equation}
I^* = I^*_1 + I^*_2 - \frac{1}{2} \log (1 - \rho^2), \label{corrGain}
\end{equation}
where $I^*_{1,2}$ is the channel capacity of the transcription factors individually, and $\rho = \frac{\Sigma_{12}}{\sqrt{\Sigma_{11}\Sigma_{22}}}$ is the noise correlation coefficient for TF concentrations. Accounting for noise correlation enhances the rate of information transmission. In fact, in the limit of perfect correlation, $\rho \to \pm 1$, the capacity is infinite. This is expected, since under the small-noise approximation and perfectly correlated noise, some combination of inputs is always noise free. Noise-free continuous variables can transmit infinite information. Fig.1 is a pictorial representation of how noise correlations are beneficial. Essentially, the information is encoded in a combination of the two inputs (such as their difference), which is subject to less noise.

\begin{figure}
   \begin{center}
      \includegraphics*[width=3.25in]{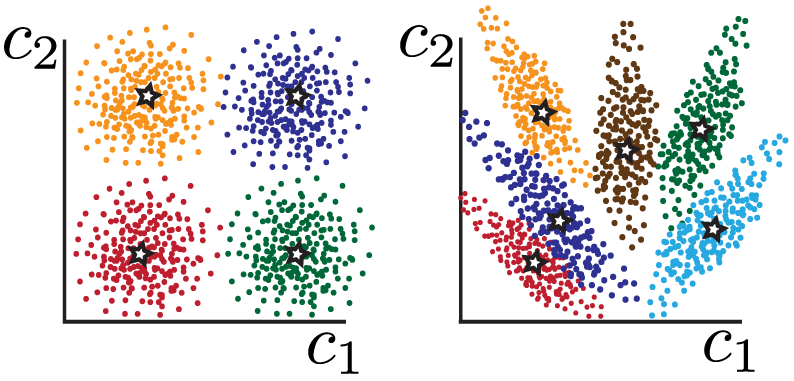}
      \caption{Benefit of correlated noise. (Left) Uncorrelated noise. Each color corresponds to a particular output response $g$. Due to noise, many inputs ($c_1$,$c_2$) correspond to the same color output. For effective signaling, the outputs (and corresponding mean inputs marked as stars) must be sufficiently spaced to avoid ambiguity. In the simple picture above, four different outputs can be reliably communicated, corresponding to two values of $c_1$ and $c_2$ which can be selected independently. (Right) The two inputs have correlated noise as reflected in the ellipsoidal scatter of input points $(c_1,c_2)$ corresponding to the same output. Six distinct outputs can be reliably communicated due to smaller spread of noise in one direction. However, the non-trivial tiling means that the six allowed values of each input can not be selected independently.}
      \label{fig:result_fig}
   \end{center}
\end{figure}

In general, the optimal input distribution (Eq.\ref{optDist}) is not separable to individual components, namely,
\begin{equation}
P^*_{TF}(c_1,c_2) \neq P^*_1(c_1)P^*_2(c_2),
\end{equation}
where $P^*_{1,2}$ is the marginal distribution for $c_{1,2}$. In a sense, $P^*_{TF}(c_1,c_2)$ is an entangled distribution, where the concentration of one TF determines the probability of observing a certain concentration of the other. Biologically, this hints at upstream interactions between the transcription factors; the form of which should be predictable from the nature of the noise correlations.

The above abstract results are not surprising. The more important question is whether noise can be correlated, i.e. $P(c_1,c_2|g) \neq P(c_1|g)P(c_2|g)$, for the physical process of binding and unbinding of multiple TFs to a promoter region. We will demonstrate this below using competing transcription factor modules.

\subsection*{Competing transcription factors}

Transcription factors regulate gene expression levels by binding to cis-regulatory regions on the DNA. The design of these regions is highly complex in both prokaryotes and eukaryotes, with overlapping TF binding sites occurring frequently \citep{Alon1,Lee,tenWolde}.

We write a simplified model of the two extremes of overlapping binding sites (Fig.2). The dominant source of noise is assumed to be the intrinsic noise from fluctuations in binding/unbinding of TF to the promoter; we address the validity of neglecting the diffusion noise in the TF concentration in the Discussion section. Details of RNAP assembly and transcription are coarse grained to a simple TF binding picture. Nonetheless, we will show that this simple model captures the essential role of noise correlations in a regulatory network. 

\begin{figure}
   \begin{center}
      \includegraphics*[width=2.75in]{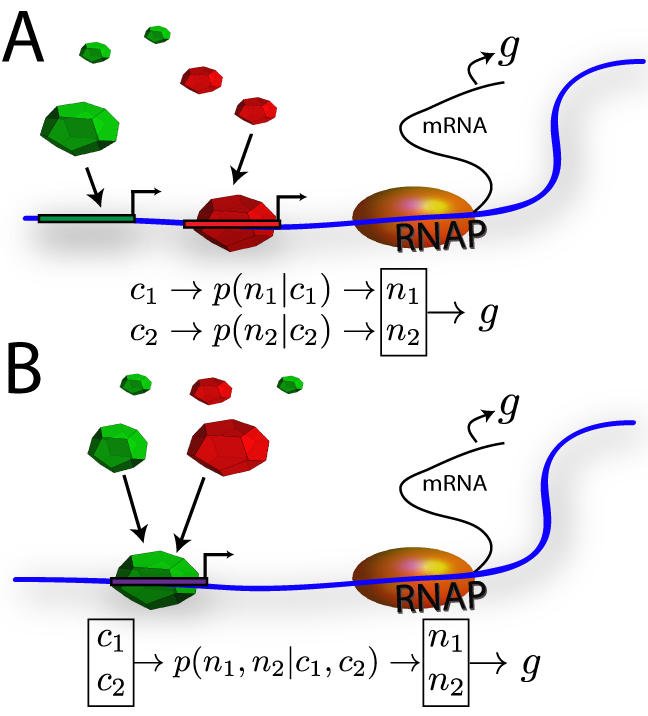}
      \caption{Independent vs. competing transcription factors. (A) Non-overlapping binding sites. Fractional binding-site occupations $n_{1,2}$ are not correlated, neither is the noise in estimates of $c_{1,2}$. The expression level $g$ is dependent on both inputs. (B) Over-lapping binding sites. $n_{1,2}$ are dependent, resulting in correlated noise in estimating $c_1$ and $c_2$. }
      \label{fig:result_fig}
   \end{center}
\end{figure}

Following the approach of \citep{BialekPNAS2}, let $n_{1,2}$ be the fractional occupation of the binding site by competing TF$_{1,2}$. $n_1 + n_2 < 1$ is the fractional occupation of the site by either TF. A binding event can occur only if the site is unoccupied, $1-n_1-n_2$ of the time.
\begin{equation}
\frac{dn_i(t)}{dt} = k c_i \big( 1 - n_1 - n_2 \big) - l n_i \ \ \ i=1,2 \label{BindingModel}
\end{equation}
The binding rate (on-rate) is proportional to the concentration of TF present, and the off-rates given by constant $l$. At thermal equilibrium these two rates, are related through the principle of detailed balance, $\frac{k c_1}{l} = \exp \big( \frac{F_1}{k_B T} )$, where $F_1$ is the free energy gain in binding for TF$_1$, with a similar expression for TF$_2$.  We rescale time so that $k=1$.

Eq.\ref{BindingModel} is a dynamical picture of the fractional occupation of the binding site by each TF. At steady state the mean fractional occupation is denoted by $\bar{n}_{1,2}$. We incorporate thermal fluctuations by introducing small fluctuations in $F_{1,2}$, capturing thermal kicks of energy that result in binding/unbinding events by effectively changing the binding energy. We do not worry about fluctuations in $c$ itself --extrinsic noise: the TF concentrations do not fluctuate; they are the fixed inputs of the system. Fluctuations in the fractional occupation of the binding site $n_{1,2}$ effectively introduce noise in the read-out of the concentrations, $p(\n|\vc)$.

With this substitution and taking the Fourier transform, the linearized fluctuations around the mean $\delta n_{1,2}$ satisfy,
\begin{eqnarray}
\begin{pmatrix}
\delta \tilde{F}_1 \\
\delta \tilde{F}_2
\end{pmatrix} = \frac{k_B T}{1 - \bar{n}_1 - \bar{n}_2}
\begin{pmatrix}
1+ \frac{-i \omega + l_1}{k_1 c_1} && 1 \\
1 && 1+\frac{-i \omega + l_2}{k_2 c_2}
\end{pmatrix}
\begin{pmatrix}
\delta \tilde{n}_1 \\
\delta \tilde{n}_2
\end{pmatrix}  \label{dFRel}
\end{eqnarray}
where tilde denotes the Fourier-transform, $\delta\tilde{n}_1(\omega) = \int_0^\infty \delta n(t) e^{i\omega t} dt$. In vectorial form, the relation becomes, $\delta \tilde{\bf F} = \Lambda \delta \tilde{\bf n}$.

Eq.\ref{dFRel} relates incremental fluctuations in read-out $\delta \n$ with fluctuations in free energy $\delta \F$. This is a linear response relation, with the free energy playing the role of the driving force (for details, see \citep{BialekPNAS2}). Using fluctuation dissipation theorem \citep{Kubo}, we calculate the power-spectrum of noise in $\n$.
\begin{equation}
S_n(\omega) = \frac{2k_BT}{\omega} \Im(\Lambda^{-1}), \label{PowerSpectrum}
\end{equation}
with $\Im$ denoting the imaginary part. From $S$, we can compute the covariance matrix,
\begin{equation}
\langle \delta\n^T \delta\n \rangle = \int_{-1/\tau_{int}}^{1/\tau_{int}} \frac{d\omega}{2\pi} S_n(\omega). \label{intTime}
\end{equation}
$\tau_{int}$ denotes the integration time of the site. For now, we assume $1/\tau_{int} \to \infty$ to compute the instantaneous fluctuations in the binding read-out. Later, we will consider biologically relevant integration times. With proper normalization, we can compute the correlation coefficient (Fig. 3A). The correlation-coefficient is negative, since a more than expected occupation of the site by one TF will clearly result in less than expected occupation by the other.

\begin{figure}
   \begin{center}
      \includegraphics*[width=3.75in]{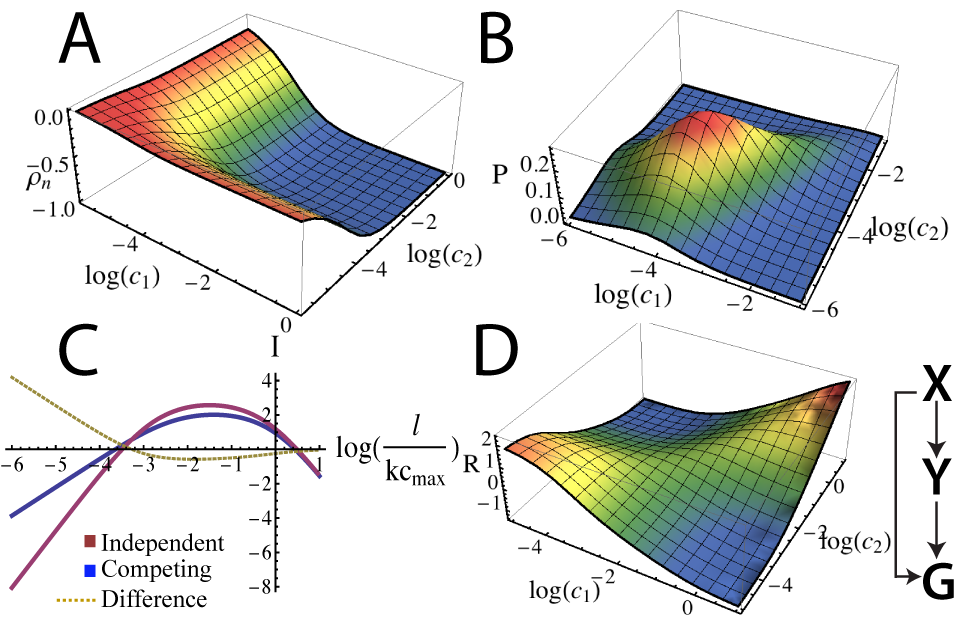}
      \caption{Competing transcription factors. (A) Correlation coefficient of readouts $n_1$ and $n_2$, for $l=10^{-4}$ and $c_{min}=10^{-3}$ as a function of log input TF concentration. At high concentrations, a higher than expected readout of one TF implies a lower than expected readout of the other, resulting in a negative correlation coefficient. (B) The optimal input distribution for the same parameter values. (C) The channel capacity in bits for the interacting and non-interacting case of two TFs (blue and red curves respectively) as a function of logarithm of rescaled $l$. The y-offset is arbitrary. Dashed curve denotes their difference. At biologically relevant $l=10^{-4}$, interacting TFs have higher channel capacity. (D) The log likelihood of observing TF concentration $(c_1,c_2)$ compared to what is expected from independent distributions. It is much more likely to observe both TFs at either low or high concentrations together. This suggests that one TF positively regulates the other, feed-forward motif (right).}
      \label{fig:result_fig}
   \end{center}
\end{figure}

Finally, we need to relate the noise in $\delta\n$ to the noise in the estimated TF concentrations. To do so, we account for the sensitivity of $\n$ to the TF concentrations. For example, a very large $c_1$ results in $n_1 = 1$ with little noise. This read-out, however, is not very sensitive to changes in $c_1$, and not useful in detecting concentration changes. Define the matrix, $\Omega_{ij} =  \frac{\partial c_i}{\partial \bar{n}_j}$. The covariance matrix for the noise in TF concentrations is given by,  $\Sigma = \langle \delta\vc^T \delta\vc \rangle = \Omega \langle \delta\n^T \delta\n \rangle \Omega^T$.

In equating the covariance matrix in TF concentration to $\Sigma$ (covariance matrix for a fixed $g$) of the previous section, we have introduced the extra assumption that the dominant noise in the channel going from $\vc$ to $g$ is from the binding noise and not the expression-level. Noise in $g$ is assumed negligible and need not be propagated backwards and included in $\Sigma$. Since noise in $g$ is most commonly shot-noise \citep{Berg}, this assumption is reasonable when expression-levels are high. This also means that our results will not depend on the functional form of $g$ on $\vc$ (for the case when they do for one input see \citep{BialekPRE1,BialekPRE2,BialekPRE3}). We will relax this assumption below for the numerical simulations of the feed-forward loop.

We compute the optimal joint-distribution of input concentration $P^*_{TF}(c_1,c_2)$ by plugging the covariance matrix in Eq.\ref{optDist} (Fig. 3B). Moreover, Eq.\ref{chanCap} tells us the channel capacity, or the maximum information transmission rate. Fig. 3C plots channel capacity of two interacting TFs and two independent ones as a function of logarithm of off-rate $\log_{10} (l)$. The interacting TFs have a higher channel capacity in the biologically relevant regime where $l \sim 10^{-4}$ and $c_{min} \sim 10^{-3}$ (see below). This result does not depend sensitively on the lower bound of the TF concentration $c_{min}$; in fact, channel capacity is finite even when $c_{min} = 0$. The lower bound is enforced to ensure a minimum of one signaling molecule in the cell. The channel capacity has not increased simply because more signaling molecules are used in the interacting case. In fact, at these parameters, the mean input TF concentration, $\int \int dc_1 dc_2 (c_1 + c_2) P^*_{TF}(c_1,c_2)$, is approximately $30 \%$ less than that of the non-interacting channel.

The optimal joint-distribution of input concentrations (Fig. 3B) is `entangled' and no longer separable, $P^*_{TF}(c_1,c_2) \neq P^*_1(c_1)P^*_2(c_2)$. With an entangled distribution the system can explore degrees of freedom not present with two independent input distributions. In Fig. 3D, we plot the log likelihood of observing joint concentration $(c_1,c_2)$ compared to observing $c_1$ and $c_2$ independently from their marginal distributions. $R = \log_{10} \frac{P^*_{TF}(c_1,c_2)}{P^*_1(c_1)P^*_2(c_2)}$, where $P^*_1(c_1) = \int dc_2 P^*_{TF}(c_1,c_2)$ is the marginal distribution of TF$_{1}$, with a similar expression for TF$_2$.

Fig. 3D implies that the two TFs are no longer passive and in fact cross-regulate each other. It is $\sim$10 times less likely to observe one TF at a high concentration and the other at a low concentration simultaneously, compared to what is expected if they were independent. Similarly, it is $\sim$10 times more likely to observe high concentrations of one TF if the concentration of the other is also high. This suggests that one TF positively regulates the other (feed-forward motif in Fig. 3D). 


Where does a biological system lie in the abstract parameter space sketched above? As noted, we have rescaled time so that $k=1$, and measured concentration in units of $c_{max} = 1$. The only parameters left are the off-rate $l$ and $c_{min}$. In a real cell, we expect a maximum of roughly 1000 TF molecules (or a dynamic range of 1-1000 TF molecules) in a volume of $\sim 1 \ \mu m^3$ \citep{Hwa}. Hence, the minimum allowed concentration is $c_{min} = 10^{-3}$. A typical equilibrium constant of TF binding to DNA is $K_{eq} \sim 10^{10}\ M^{-1}$ \citep{Phillips}. Putting all this together, we find $l \sim 10^{-4}$. It is possible then that a real biological regulatory system can transmit more information by incorporating overlapping binding sites and an upstream positive regulation between the TFs.

\subsection*{Integration time and cooperativity}

To compute the channel capacity above, we used the instantaneous variance of the binding-site fractional occupation, $\langle \delta n(t)^2 \rangle$. In reality, however, a cell will integrate the occupation of the binding site for some time. The general theory proposed above is also valid in the limit $\tau_{int} \to \infty$. A longer integration time typically decreases both the variance and covariance by a factor $1/\tau_{int}$; the correlation coefficient $\rho$ is unaffected. An entangled joint-distribution of inputs is in general still more optimal than a separable one. However, the specific form of the frequency-dependance of the noise can make the role of integration time more complicated. We examine information transmission in the above model for biologically relevant integration times.

The binding of a TF is a binary --on/off-- signal for transcription of mRNA. The notion of a fractional occupation inherently assumes averaging over a series of binding and unbinding events. The first stage of integration is through transcription: the amount of time a TF is bound to DNA is approximately proportional to the amount of mRNA transcribed. Therefore, the life-time of the mRNA $\tau_e$ sets the transcriptional integration time-scale. If mRNA lifetime is long, more mRNA molecules accumulate, resulting in a more precise value of the average time that TF was bound. mRNAs in turn translate into protein; accumulation of proteins --with lifetime $\tau_g$-- is the second stage of integration. 

Although generally $\tau_g \gg \tau_e$ \citep{Swain08}, translation is a discontinuous process, with other sources of interruptions besides binding fluctuations --i.e. chromatin remodeling and mRNA splicing in eukaryotes \cite{Kaern05} and transcriptional bursting in prokaryotes \cite{Golding05}. Naively, transcriptional integration removes fluctuations with frequencies higher than $\tau_e^{-1}$; translational integration has no frequency dependance --because it is punctuated-- and simply reduces the variance of fluctuations by a factor $1/\tau_g$. After the integration, the binding noise is estimated as, 

\begin{equation}
\langle \delta\n^T \delta\n \rangle \approx \frac{1}{\tau_g} \int_{-1/\tau_e}^{1/\tau_e} \frac{d\omega}{2\pi} S_n(\omega). \label{intTime2}
\end{equation}

The power spectrum for interacting and non-interacting TFs is shown in Fig. 4A from analytical calculations (Eq.\ref{PowerSpectrum}) and numerical simulations of Eq.\ref{BindingModel} (Methods). For a non-interacting TF, assuming $kc \gg l$, $n$ fluctuates on the timescale $(kc)^{-1}$ (see derivation in \citep{BialekPNAS2}). Surprisingly, interacting TFs also show fluctuations at the slower timescale $l^{-1}$; refer to Fig. 4B and the power spectrum in Fig. 4C. However, the long wavelength fluctuations are almost perfectly anti-correlated between the two read-outs $n_1$ and $n_2$ ($\rho \to -1$); when the read-outs are combined the remaining fluctuations have timescale $(kc)^{-1}$. Since the power-spectrum of the competing TFs has a narrower width $~l$, the integration time must be longer than $l^{-1}$ for the noise to change substantially. TF dissociation rates can be slow, $l \sim 10^{-3} \ s^{-1}$ \citep{Ong,Phillips}. A typical mRNA lifetime of minutes $\tau_e \sim 10^2 \ s$ averages out the fast fluctuations at rate $k \sim 1 s$ (assuming a typical TF concentration of $100 \ nm$) but does not filter the low-frequency fluctuations at rate $l$. The long protein lifetime $\tau_g$ --typically many minutes to hours \citep{Swain08}-- averages over many binding and unbinding events.

\begin{figure}
   \begin{center}
      \includegraphics*[width=3.75in]{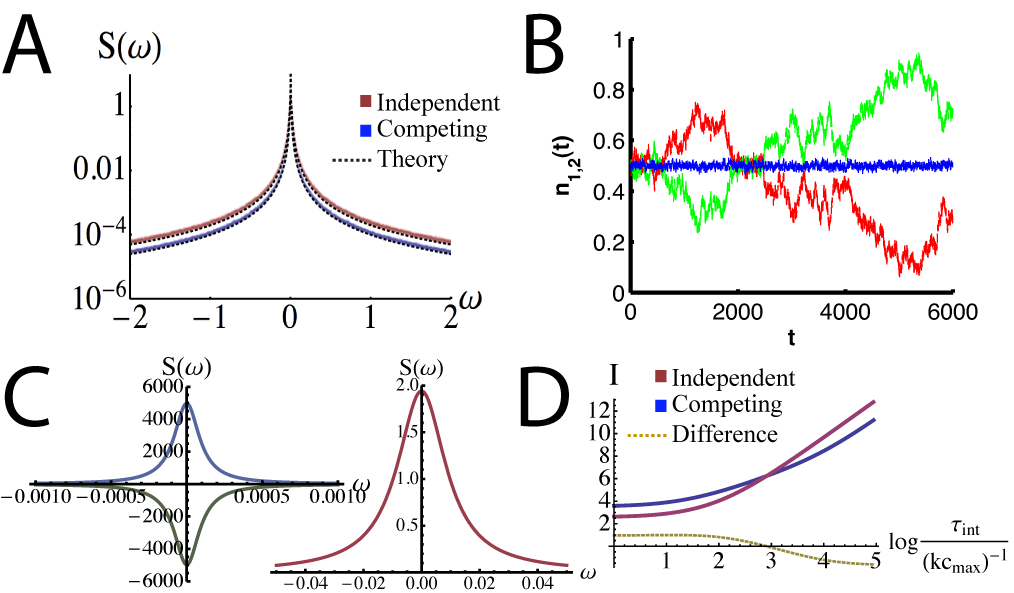}
      \caption{Integration time. (A) Power spectrum of the fluctuations in fractional binding $n$ for both competing and independent TFs. The analytical calculation is in good agreement with simulation results (shown here for $k=1$, $c=0.01$, and $l=10^{-4}$; methods). (B) A typical time-series of fluctuations in $n_{1,2}$ --red and green curves-- exhibiting short-wavelength fluctuations (time-scale $(kc)^{-1}$) and long-wavelength fluctuations (time-scale $l^{-1}$). The long-term fluctuations are almost perfectly anti-correlated and can be removed by averaging the two inputs --blue curve. (C) Analytical power spectrum of the competing (left) and independent (right) TFs. Cross-power spectral density is also plotted for the competing case, which is negative because of the anti-correlations. The width of the power spectrum is order of $kc$ for independent TFs but much narrower, order of $l$, for competing TFs. (D) Channel capacity as a function of base-10 logarithm of normalized integration time for both types of TFs. Competing TFs outperform independent TFs for integration times up to $\sim10^3(kc_{max})^{-1}$.}
      \label{fig:result_fig}
   \end{center}
\end{figure}

We explicitly compute the chancel capacity for competing and independent TFs as a function of the integration time (Fig. 4D). The correlation coefficient of the fluctuations between the read-outs does not diminish with increasing integration time. For low dissociation rates, $kc_{max}/l = 10^4$, competing TFs transmit more information than independents TFs up to an integration time $10^3/kc_{max}$ --roughly $\sim 10^2 \ s$ using above parameters, which is comparable to the biologically relevant integration time set by mRNA lifetime. We also explored the impact of cooperative binding of TFs by adding a Hill coefficient $\lambda$ to the concentrations in Eq.\ref{BindingModel} ($c_{1,2} \to c_{1,2}^\lambda$). The relative advantage of competing TFs disappears with increasing cooperativity. When $\lambda = 5$, the advantage of competing TFs disappears for any biologically relevant integration time. For $\lambda = 0.5$, however, the channel capacity is higher with competition --a lager increase compared with the uncooperative case-- and persists to arbitrary large integration times, $\tau_{int} \to \infty$. It is conceivable that competing TFs may transmit more information than independent TFs in the limit of low dissociation rates or negative cooperativity for biologically relevant integration times; see the Discussion section for the importance of diffusion noise and its connection to cooperativity.

\subsection*{Feed-forward motif}

The fact that interacting TFs have correlated noise is not surprising. The entangled optimal input distribution calculated above implied that one TF positively regulated the other upstream --reminiscent of a feed-forward motif. Naturally, the question arises if a realistic biological model of feed-forward gene regulation can take advantage of correlated noise in the inputs. Can dynamical joint repression/activation of a target gene encode the signal in a combination of the two correlated inputs which is subject to less noise? Is it possible to optimize the input distribution using realistic gene regulatory modules? We answer these questions by numerically computing channel capacity of a feed-forward loop (FFL), where up-stream one TF regulates the other, and downstream both jointly regulate the expression level of the target gene. Another purpose of the numerical approach is to relax the restrictive assumptions required for the above analytical derivations, in particular the small-noise approximation, Gaussian form of the noise, one-to-one correspondence between input TF concentrations and the output expression level, and negligible output noise.

In the following analysis we neglect the details of how fluctuations in read-out of TF concentrations become correlated --one mechanism is overlapping binding sites (see above)-- and simply introduce a general phenomenological model of input noise (Methods). The model of competing TFs considered above closely resembles FFL Type 2 (+++OR; Fig. 5). Although many microscopic mechanisms can potentially generate correlated noise (interference), upstream cross-regulation of transcription factors is limited to certain well-characterized gene-regulatory modules. Our purpose is to first confirm that typical gene regulatory networks can generate a close to optimal entangled distribution, and second, to check if Hill-type regulatory logic can combine correlated input noise --for example add anti-correlated inputs-- to maximize information transmission.

\begin{figure}
   \begin{center}
      \includegraphics*[width=6.2in]{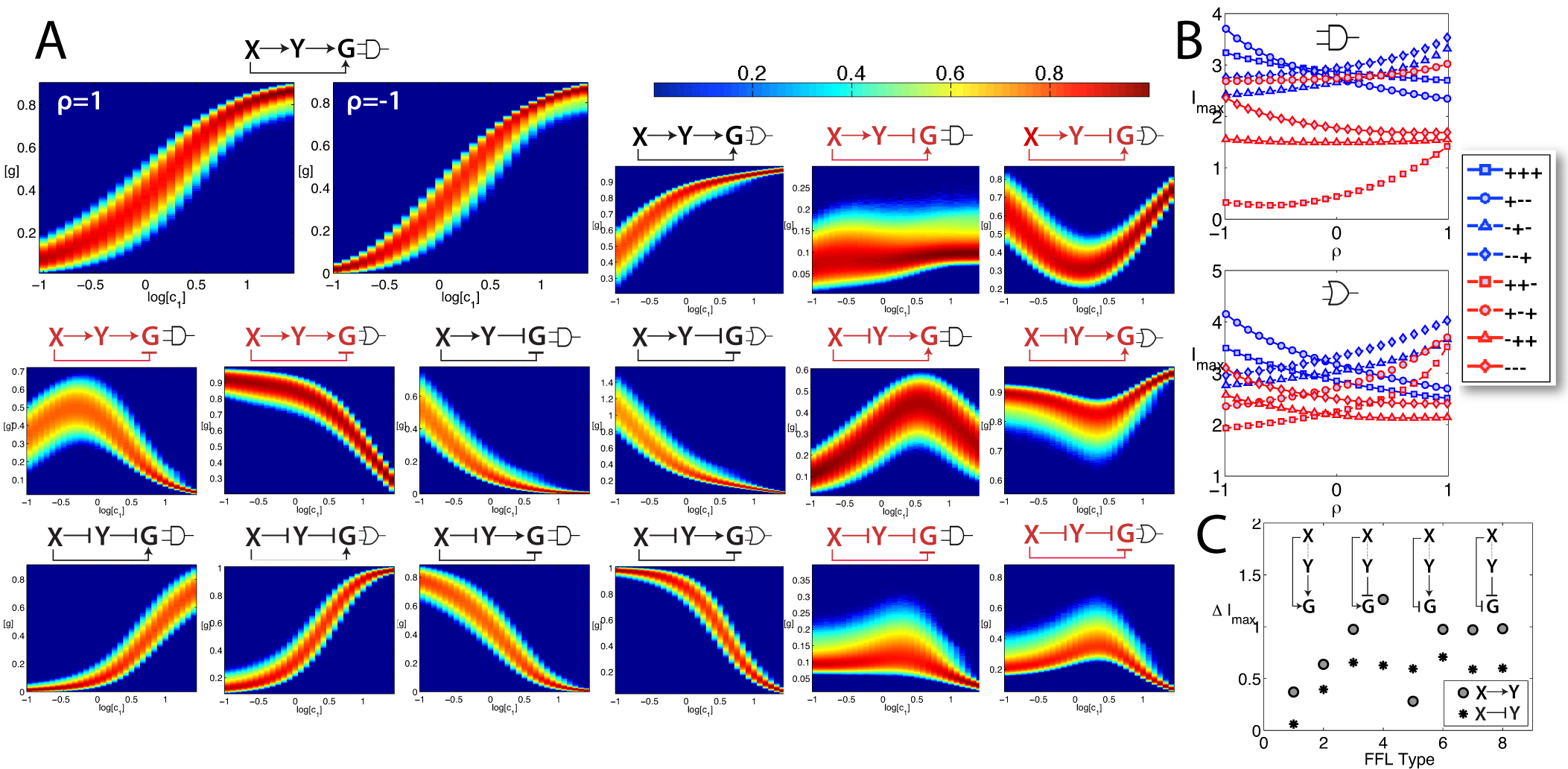}
      \caption{Numerical simulation of the feed-forward motif. There are 16 types of FFLs. The 8 incoherent networks are denoted in red. (A) Probability of output expression level $g$ as a function of log input X concentration $c_1$, $p(g|c_1)$, normalized to the displayed colorbar. Incoherent networks can exhibit non-monotonic response. The larger figures show the response for Type 1 coherent FFL when X-Y noise correlation coefficient $\rho=1$ and $\rho=-1$. Anti-correlated input noise results in reduced uncertainty in response $g$ and a higher channel capacity. The smaller figures have $\rho=0$. (B) Channel capacity (bits) for all FFLs as a function of input noise correlation coefficient $\rho$ (AND-gate networks top, OR-gate networks bottom). The legend denotes the sign (+ activator, - repressor) of X-Y, X-G, and Y-G regulation respectively. The incoherent networks (red curves) have generally lower channel capacity than the coherent ones (blue curves). Networks where X and Y regulate G with the same sign (solid curves) have enhanced channel capacity for negative $\rho$, those with opposite sign (dashed curves) improve with positive $\rho$. (C) For each network type, the circle (star) denotes maximum gain in capacity from noise correlations for X activating (repressing) Y. Odd and even network numbers correspond to AND and OR gates respectively. The upstream cross-regulation between X and Y is important in taking advantage of noise correlations.}
      \label{fig:result_fig}
   \end{center}
\end{figure}

In the feed-forward motif, input TF $X$ regulates TF $Y$, and both jointly regulate the expression level $g$ of the target gene. Since each regulatory function can be either an activator (positive) or a repressor (negative), there are eights types of FFLs \citep{Alon2}. If the sign of the direct regulation of $g$ by $X$ is the same as the sign of regulation of $g$ by $X$ through $Y$ (sign of $X$ to $Y$ regulation multiplied by that of $Y$ to $g$) then the network is called coherent. The four networks that are not coherent are called incoherent. Moreover, if both $X$ and $Y$ are required to express $g$, the FFL is designated with an AND-gate. If either TF can result in expression, an OR-gate designation is used. Including the gate, there are sixteen unique types of FFL (Fig. 5). 

We have systematically simulated the stochastic dynamics of all FFL types for a range of input TF concentrations (see Methods). We focus on the intrinsic noise in the joint-regulation of $g$ by TFs $X$ and $Y$ which has two sources: output noise due to stochastic synthesis and degradation of $g$, and input noise in read-outs of TF concentrations. The input noise of TF $X$ is correlated with that of $Y$ with correlation coefficient $\rho$. The extrinsic contribution of noise from the upstream regulation of $Y$ by $X$ is considered negligible --its inclusion does not qualitatively change our results. The cross-regulations sets up the joint-distribution of $X$ and $Y$, $P(c_1,c_2)$, for a given input distribution of $X$, $P_{in}(c_1)$.

The probability of observing  expression level $g$ for input concentration $c_1$, $P(g|c_1)$, is computed by sampling the steady state expression levels for many runs. As evident in Fig. 5A, the noise distribution $P(g|c_1)$ is not Gaussian in general; non-linearities in the model result in lopsided distributions. Furthermore, the incoherent FFLs are non-monotonic functions of input $c_1$ to $g$. The same expression level $g$ can correspond to more than one intended input; also in contrast to the earlier assumptions.

We have computed the channel capacity of each FFL by numerically optimization mutual information between output distribution, $p(g) = \int P(g|c_1)P_{in}(c_1)dc_1$ and TF $X$ input distribution, $P_{in}(c_1)$ over all input distributions (see Methods). Coherent FFLs have generally a higher capacity than the incoherent ones; coherent AND loops on average can transmit 2.8 bits vs 1.6 bits for incoherent ANDs, for OR networks the capacity is 3.1 bits vs 2.4 bits. The non-monotonicity of the incoherent networks creates ambiguities in mapping the output to the intended input. Correlations in fluctuations of read-outs of $X$ and $Y$ concentrations increase channel capacity in all FFLs (Fig. 5B). Networks where $X$ and $Y$ regulate $g$ with the same sign (both activate or both repress $g$) enhance their channel capacity when the input noise correlation coefficient, $\rho$, is negative; networks with opposite signs improve with positive $\rho$. This is expected since when $X$ and $Y$ regulate $g$ in the same way, inputs are effectively added; adding two channels with anti-correlated noise reduces the noise. Similarly, subtracting inputs with correlated noise results in noise reduction. For our choice of parameters, we observe for example a 13\% increase in channel capacity of Type 1 Coherent FFL, +++AND (sign of $X$-$Y$,$X$-$G$, and $Y$-$G$ regulation respectively), when $\rho=-1$ compared to $\rho=0$; coherent FFL +-{-}OR --which roughly correspond to our earlier model of TFs competing for the same binding site-- showed an improvement of 31\%. Other choices of parameters produced similar results.

We claimed above that interacting TFs with an optimal input joint-distribution outperformed non-interacting TFs when noise correlations were incorporated despite the increase in the noise of the individual channels from competition. This observation broadly holds in our simulations. For example, the highest capacity network, FFL -{-}+OR, has a channel capacity of 4.0 bits with an input noise correlation coefficient $\rho=1$; whereas the same network with the variance of input noise reduced by a factor two ($q=0.5$, see Methods) and no correlation has a capacity of 3.7 bits. Incorporating correlations at the expense of higher noise variance seems to be a beneficial strategy. Lastly, we stressed the importance of up-stream cross-regulation between the two TFs as means of constructing the optimal entangled joint-distribution. To confirm that the gain in capacity from noise correlations is not simply due to a reduction in noise by a clever addition/subtraction of the inputs at the module regulating $g$, we compared the maximum gain in capacity from correlations for each network to its `sister' network (sign of $X$-$Y$ regulation flipped). Fig. 5C shows that the upstream $X$-$Y$ regulation is instrumental in determining the gain; whether $X$ activates or represses $Y$ further optimizes the input joint-distribution and in turn the channel capacity.

\section*{Discussion}

We showed that quite generally a signaling pathway with interference --correlations in input noise due to microscopic interactions of the signaling molecules-- can optimize information transmission by implementing cross-talk upstream between the interacting molecules --such that concentration of one input depends on the other.

Concentration-dependent transcriptional regulation is particularly important at the developmental stage. Concentration of morphogens dictate cell fate, for example, resulting in patterning of the {\it Drosophila} embryo along the dorsoventral  axis \citep{Anderson}. It is likely that the embryo has optimized information transmission to ensure accurate patterning and later development. Gene regulation using a combination of transcription factors is also a common theme in development \citep{Davidson}.

Xu et al. \citep{Rushlow} have observed the feed-forward motif in regulation of gene {\it Race} in the {\it Drosophila} embryo. They report that intracellular protein Smads sets the expression level of {\it zerkn$\ddot{u}$llt} (zen), then Smads in combination with zen (two-fold input) directly activate {\it Race}. Analysis of binding site of Smads and zen reveals slight overlaps, and experiments indicate that one protein facilitates binding of the other to the enhancer. This interaction can result in a similar positive correlation coefficient in TF concentration estimates derived above. The previously proposed suggestion \citep{Rushlow} that feed-forward motif increases sensitivity to the input signal does not explain why the target is regulated by both the initial input and the target transcription factor. Proposed dynamical features associated with the feed-forward loop \citep{Alon1,Alon2} do not explain the need for overlapping binding sites and TF interactions at binding. 

Another example of a feed-forward motif coupled to binding interactions is the joint-regulation of {\it even-skipped} (eve) stripe 2 by {\it bicoid} (bcd) and {\it hunchback} (hb). Small et al. \citep{Levine1} report cooperative binding interactions between bcd and hb and a clustering of their binding sites in the promoter region. Upstream, bcd positively regulates transcription of hb. Similarly, Ip et al. \citep{Levine2} have observed joint-activation of gene {\it snail} (sna) by {\it twist} (twi) and {\it dorsal} (dl), which also exhibit cooperative binding interactions. dl directly regulates transcription of twi upstream.

More generally, other forms of cross-talk besides transcriptional regulation can be used. For instance, in regulation of anaerobic respiration in {\it E. coli}, regulators NarP and NarL are jointly-regulated through phosphorylation by histidine kinase NarQ. NarL is also phosphorylated by kinase NarX. Downstream, NarP and NarL share the same DNA binding site \citep{Stewart}. It is not, however, clear if optimizing channel capacity is relevant for this system. It is also possible that interference is implemented using other schemes than DNA binding, for example, through cooperative interactions of signaling molecules with scaffold proteins \citep{Wendell}. 

We have shown that TF interaction at overlapping binding sites plus upstream cross-regulation can enhance information transmission compared to non-interacting TFs. This is consistent with the frequent observation of the feed-forward motif ending in overlapping binding sites in developmental gene networks.  Although the feed-forward motif has been proposed before for optimizing information transmission in regulatory networks \citep{BialekPRE2}, we emphasize that our approach is fundamentally different, since it stems from correlated binding noise, and physically requires existence of TF interactions at the binding level. This is indeed what is experimentally observed in the three examples discussed above. Diamond motifs, where inputs are transmitted independently and then recombined later, have also been proposed as mechanisms of increasing gain in signaling pathways \citep{Ronde12}.

The key assumptions in the above model were as follows: primary source of noise is intrinsic input noise from read-out of TF concentrations, and negligible extrinsic noise from cross-regulation. Inclusion of extrinsic noise simply reduces the overall channel capacity and does not modify the relation between input noise correlations and upstream cross-regulation. For the case of TFs competing for the same binding site, the intrinsic noise was assumed to be dominated by binding fluctuations as opposed to diffusion noise. This assumption is valid in the limit of low dissociation rates and low cooperativity \cite{Tkacik08}. Moreover, this limit can be potentially consistent with biologically relevant integration times: competing TFs with low dissociation rates have a higher channel capacity than independent TFs even when integration time is comparable to typical mRNA lifetime; with negative cooperativity, the integration time can be arbitrarily large. Even if noise is dominated by diffusion, other mechanisms --such as di- or multi-merization of the signaling molecules, or cooperative active transport-- may generate correlations in diffusion noise. The same framework can then connect multimerizaiton of signaling molecules to their upstream cross-regulation. Although our example focussed on the particular case of competing TFs, we stress that in general any signaling pathway with correlated noise can transmit more information when optimized with cross-talk between the inputs.

A myriad of logical regulatory circuits have been proposed through use of overlapping binding sites and interacting TFs \citep{tenWolde,Hwa}. It is worthwhile to see if the upstream TF regulatory network of these systems can be correctly predicted from TF binding-site overlap or other interactions using the methodology outlined above (Eq.\ref{optDist}). Such analysis requires knowledge of the input noise, which can be obtained by a bioinformatics approach, where the binding sequence of each TF is examined for overlap, or direct measurements of the noise using single-molecule techniques \citep{Ong} for other types of interactions.

\section*{Acknowledgements}

The author thanks Boris Shraiman for helpful discussions and critical reading of the manuscript, and Bill Bialek for introduction to the subject. This research was supported in part by the National Science Foundation under Grant No. NSF PHY11-25915.


\begin{thebibliography}{10}

\bibitem[Wolpert et al, 2006]{Wolpert1} Wolpert, L., et al. 2006. Principles of Development (Oxford Univ Press), 3rd Ed.

\bibitem[Wolpert, 1969]{Wolpert2} Wolpert, L. 1969. Positional information and the spatial pattern of cellular differentiation. J Theor Biol {\bf 25}:1-47.

\bibitem[Small et al, 1992]{Levine1} Small, S., A. Blair, and M. Levine. 1992. Regulation of even-skipped stripe 2 in the Drosophila embryo. {\it The EMBO Journal} {\bf11}:4047-4057.

\bibitem[Ip et al, 1992]{Levine2} Ip, Y. T., R. E.  Park, D. Kosman, K. Yazdanbakhsh, and M. Levine. 1992. dorsal-twist interactions establish snail expression in the presumptive mesoderm of the Drosophila embryo. Genes Dev {\bf 6}:1518-1530.

\bibitem[Elowitz et al, 2002]{Elowitz1} Elowitz, M. B., A. J. Levine, and E. D. Siggia, P. S. Swain. 2002. Stochastic gene expression in a single cell. Science {\bf 297}:1183-1186.

\bibitem[Swain et al, 2002]{Elowitz2} Swain, P. S., M. B. Elowitz, and E. D. Siggia. 2002. Intrinsic and extrinsic contributions to stochasticity in gene expression. Proc Natl Acad Sci USA {\bf 99}:12795-12800.

\bibitem[Paulsson, 2004]{Paulsson} Paulsson, J. 2004. Summing up the noise in gene networks. Nature {\bf 427}:415-418. 

\bibitem[Tckacik \& Walczak, 2011]{WalczakReview} Tkacik, G., and A. M. Walczak 2011. Information transmission in genetic regulatory networks: a review. J. Phys.: Condens. Matter {\bf 23}:153102.

\bibitem[Gregor et al, 2007]{Gregor} Gregor, T., D. W. Tank, E. F. Wieschaus, and W. Bialek. 2007. Probing the limits to positional information. Cell {\bf 130}:153-164. 

\bibitem[Shannon, 1949]{Shannon} Shannon, C. E. 1949. Communication in the presence of noise. Proc IRE {\bf 37}:10-21.

\bibitem[Cover \& Thomas, 1991]{Cover} Cover, T. M., and J. A. Thomas. 1991. Elements of Information Theory (Wiley, New York).

\bibitem[Tkacik et al, 2008]{BialekPNAS} Tkacik, G., C. G. Callan, and W. Bialek. 2008. Information flow and optimization in transcriptional regulation. Proc Natl Acad Sci USA {\bf 105}:12265-12270.

\bibitem[Higham, 2001]{Higham} Higham, D. J. 2001. An algorithmic introduction to numerical simulation of stochastic differential equations. SIAM Rev. {\bf 43}:525-46.

\bibitem[Papoulis, 1991]{Papoulis} Papoulis, A. 1991. Probability, Random Variables, and Stochastic Processes (McGrawÐHill, New York).

\bibitem[Shen-Orr et al, 2002]{Alon1} Shen-Orr, S. S., R. Milo, S. Mangan, and U. Alon. 2002. Network motifs in the transcriptional regulation network of Escherichia coli. Nat Genet {\bf 31}:64-68.

\bibitem[Mangan \& Alon, 2003]{Alon2} Mangan, S., and U. Alon. 2003. Structure and function of the feed-forward loop network motif. Proc Natl Acad Sci USA {\bf 100}:11980-11985.

\bibitem[Tyson et al, 2003]{Tyson} Tyson, J. J., K. C. Chen, B. Novak. 2003. Sniffers, buzzers, toggles and blinkers: dynamics of regulatory and signaling pathways in the cell. Curr. Opin. Cell Biol. {\bf 15}, 221-231.


\bibitem[Gregor et al, 2005]{Gregor2} Gregor, T., W. Bialek, R. R. de Ruyter van Steveninck, D. W. Tank, and E. F. Wieschaus. 2005. Diffusion and scaling during early embryonic pattern formation. Proc Natl Acad Sci USA 102:18403Ð18407.

\bibitem[Powell, 1978]{Powell} Powell, M. J. D. 1978. A Fast Algorithm for Nonlinearly Constrained Optimization Calculations, Numerical Analysis, ed. G.A. Watson, Lecture Notes in Mathematics, Springer Verlag, Vol. 630. 


\bibitem[Lee et al, 2002]{Lee} Lee, T. I., et al. 2002. Transcriptional regulatory networks in Saccharomyces cerevisiae. Science {\bf 298}:799-804.


\bibitem[Tkacik et al, 2008]{BialekPRE1} Tkacik, G., C. G. Callan, and W. Bialek. 2008. Information capacity of genetic regulatory elements. Phys Rev E {\bf 78}:11910. 


\bibitem[Little et al, 2011]{Little} Little, S. C., G. Tkacik, T. B. Kneeland, E. F. Wieschaus, and T. Gregor. 2011. The Formation of the Bicoid Morphogen Gradient Requires Protein Movement from Anteriorly Localized mRNA. PLoS Biol 9(3): e1000596.


\bibitem[Raser \& O'Shea, 2004]{O'Shea}Raser, J. M., E. K. O'Shea. 2004. Control of stochasticity in eukaryotic gene expression. Science {\bf 304}:1811-1814.

\bibitem[Newman et al, 2006]{Newman} Newman, J. R., et al. 2006. Single-cell proteomic analysis of S. cerevisiae reveals the architecture of biological noise. Nature {\bf 441}:840-846.

\bibitem[Rosenfeld et al, 2005]{Rosenfeld} Rosenfeld, N., et al. 2005. Gene regulation at the single-cell level. Science {\bf 307}:1962-1965.



\bibitem[Hermsen et al, 2006]{tenWolde} Hermsen, R., S. Tans, P. R. tenWolde. 2006. Transcriptional Regulation by Competing Transcription Factor Modules. {\it PLoS Comp. Biol.} {\bf 2}:e164.

\bibitem[Bialek \& Setayeshgar, 2005]{BialekPNAS2} Bialek, W., and S. Setayeshgar. 2005. Physical limits to biochemical signaling. Proc Natl Acad Sci USA {\bf 102}:10040-10045.

\bibitem[Kubo, 1966]{Kubo} Kubo, R. 1966. The fluctuation-dissipation theorem. Rep Prog Phys {\bf 29}:255-284.

\bibitem[Berg \& Purcell, 1977]{Berg} Berg, H. C., and E. M. Purcell. 1977. Physics of chemoreception. Biophys J {\bf 20}:193-219.

\bibitem[Walczak et al, 2010]{BialekPRE2} Walczak, A. M., G. Tkacik, and W. Bialek. 2010. Optimizing information flow in small genetic networks. II. Feed-forward interactions. Phys Rev E {\bf 81}:041905.

\bibitem[Tkacik et al, 2011]{BialekPRE3} Tkacik, G., A. M. Walczak, and W. Bialek. 2011. Optimizing information flow in small genetic networks. III. A self-interacting gene. Phys Rev E {\bf 85}:041903.

\bibitem[de Ronde et al, 2012]{Ronde12} de Ronde, W. H., F. Tostevin, and P. R. ten Wolde 2012. Feed-forward loops and diamond motifs lead to tunable transmission of information in the frequency domain. Phys Rev E {\bf 86}:021913.

\bibitem[Tkacik et al, 2008]{Tkacik08} Tkacik, G., T. Gregor, and W. Bialek 2008. The role of input noise in transcriptional regulation. PLoS ONE {\bf 3}: e2774.

\bibitem[Buchler et al, 2003]{Hwa} Buchler, N. E., U. Gerland, and T. Hwa. 2003. On schemes of combinatorial transcription logic. Proc Natl Acad Sci USA {\bf 100}:5136-5141.

\bibitem[Bintu et al, 2005]{Phillips} Bintu, L., et al. 2005. Transcriptional regulation by the numbers: Applications. Curr Opin Genet Dev {\bf 15}:125-135.

\bibitem[Shahrezaei \& Swain, 2008]{Swain08} Shahrezaei, V., and P.S. Swain 2008. Analytical distributions for stochastic gene expression. Proc Natl Acad Sci USA {\bf 105}:17256-17261.

\bibitem[Kaern et al, 2005]{Kaern05} Kaern, M., T. C. Elston, W. J. Blake, and J. J. Collins 2005. Stochasticity in gene expression: from theories to phenotypes. Nat. Rev. Genet. {\bf 6}:451-464.

\bibitem[Golding et al, 2005]{Golding05} Golding, I. , J. Paulsson, S. Zawilski, and E. C. Cox (2005) Real-time kinetics of gene activity in individual bacteria. Cell {\bf 123} 1025-1036.

\bibitem[Wang et al, 2009]{Ong} Wang, Y., L. Guo, I. Golding, E. C. Cox, and N. P. Ong. 2009. Quantitative Transcription Factor Binding Kinetics at the Single-Molecule Level. Biophysical Journal Volume {\bf 96}:609-620.

\bibitem[Morissato \& Anderson, 1995]{Anderson} Morisato, D., and K. V. Anderson. 1995. Signaling pathways that establish the dorsalventral pattern of the Drosophila embryo. Annu. Rev. Genet. {\bf 29}:371-99.


\bibitem[Howard \& Davidson, 2004]{Davidson} Howard, M. L., and E. H. Davidson. 2004. cis-Regulatory control circuits in development. Dev. Biol. {\bf 271}:109-118.


\bibitem[Xu et al, 2005]{Rushlow} Xu, M.,  N. Kirov, and C. Rushlow. 2005. Peak levels of BMP in the Drosophila embryo control target genes by a feed-forward mechanism. Development {\bf 132}:1637-1647.


\bibitem[Stewart, 2003]{Stewart} Stewart, V. 2003. Biochemical Society Special Lecture. Nitrate- and nitrite-responsive sensors NarX and NarQ of proteobacteria. Biochem. Soc. Trans. {\bf 31}:1-10.

\bibitem[Good et al, 2011]{Wendell} Good, M. C., J. G. Zalatan, and W. A. Lim. 2011. Scaffold Proteins: Hubs for Controlling the Flow of Cellular Information. Science {\bf 332}:680-686.



\end{thebibliography}

\end{document}